\documentclass[12pt,a4paper]{article}
\newcommand{\be}{\begin{equation}}
\newcommand{\en}{\end{equation}}
\newcommand{\bea}{\begin{eqnarray}}
\newcommand{\ena}{\end{eqnarray}}

\oddsidemargin .7cm \textwidth 14.5cm \textheight 23cm  \voffset -1cm \topmargin 1cm \fontsize{12pt}{14pt}\selectfont
\begin{document}

\title{\textbf{Quasilocal Energy, Komar Charge and Horizon for Regular Black Holes}}

\author{Leonardo Balart \\    \\ \small CEFIMAS and  Departamento de F\'{\i}sica, \\  \small Universidad Nacional de La Plata,\\
\small C.C. 67, (1900), La Plata, Argentina\\ \\
\emph{ \small lbalart@fisica.unlp.edu.ar}}

\date{}

\maketitle

\begin{abstract}
We study the Brown-York quasilocal energy for regular black holes. We also express the identity that relates the difference of the Brown-York quasilocal energy and the Komar charge at the horizon to the total energy of the spacetime for static and spherically symmetric black hole solutions in a convenient way which permits us to understand why this identity is not satisfied when we consider nonlinear electrodynamics. However, we give a relation between quantities evaluated at the horizon and at infinity when nonlinear electrodynamics is considered. Similar relations are obtained for more general static and spherically symmetric black hole solutions which include solutions of dilaton gravity theories.
\end{abstract}

\section{Introduction}
\label{intro}

Bose and Dadhich in Ref.~\cite{Bose:1998uu} used the notion of quasilocal energy (QLE) proposed by Brown and York~\cite{Brown:1992br} and the gravitational charge defined by the Komar integral~\cite{komar} to characterize horizons of spherically symmetric static (SSS) black holes with metrics where $g_{00}=-(g_{11})^{-1}$. They obtained an identity connecting the field energy with the gravitational charge, both evaluated at the horizon and where the field energy is the function resulting from subtracting out the QLE at infinity from the total QLE contained inside a sphere of radius $r$. These authors remark that the lapse function $\sqrt{-g_{00}}$ is only necessary to calculate the gravitational charge and that the function $g_{11}$ is enough to determine the QLE. Interpreting such quantities, they indicate that the gravitational charge ``measures the strength of the gravitational pull exerted by a body" and that the gravitational field energy is related to ``the amount of curvature of space", both quantities are equals on the sphere determined by the horizon. Finally, they also remark that it is a nonvariational identity relating quantities at the horizon and at infinity, that is, in a different way to the conventional formulation of the laws of black hole mechanics, where variations of certain quantities at the horizon and at infinity are related.

On the other hand, several regular black hole solutions have been found by coupling General Relativity to nonlinear electrodynamics (see Refs.~\cite{AyonBeato:1998ub}-\cite{Bronnikov:2000vy} and references cited therein).
The regular black holes have several features because of the nonlinearities of the field equations. For example: photons propagate along null geodesics of an effective geometry depending on the nonlinear electromagnetic field, which permits light rays to travel slower that gravitational waves due to nonlinear effects~\cite{Novello:1999pg}; the thermodynamics quantities of these black holes do not satisfy the Smarr formula nor the first law~\cite{Mo:2006tb}. However, in a particular case of regular black hole, Born-Infeld theory, where the Smarr formula is not held, the first law of black hole thermodynamics is satisfied~\cite{Rasheed:1997ns}. An interesting question that can be considered here is whether the identity of Bose-Dadhich is satisfied for regular black holes. Likewise, one can ask what happens when a regular black hole with $g_{00} \neq -(g_{11})^{-1}$ is considered.

Due to the reasons mentioned above, in this paper we focus our attention on regular black holes as a particular case of SSS solutions where $g_{00}=-(g_{11})^{-1}$ and on more general SSS solutions. We first accomplish the derivation of the QLE by considering a regular black hole where the source is a nonlinear electrodynamics field. More precisely, the black hole found by Ay\'on-Beato and Garc\'{\i}a~\cite{AyonBeato:1999rg} as a purely electric solution and by Bronikov~\cite{Bronnikov:2000vy} as a purely magnetic solution. As a first step, due to the special characteristics of the nonlineal terms, one might determine whether the gravitational charge and the gravitational field energy at the horizon of the black hole are different. Afterwards one may study the identity mentioned above by giving a proof in a more convenient way, which permits us to generalize the identity when different types of SSS solutions are considered and eventually to understand the difference between the field energy and the gravitational charge at the horizon.

We can also use this procedure to obtain a similar relation at the horizon of SSS solutions where $g_{00} \neq -(g_{11})^{-1}$. Two examples of this type of black hole solutions can be given, one regular and the other do not, and thus to study whether they satisfy a similar identity as before or not. The regular solution considered is the black hole with Skyrme hair~\cite{luckock1986} resulting from gravity coupled to the Skyrme model, which is an effective meson theory where the baryons arise as topological stable fields called skyrmions. The other example is a solution to low energy string theory representing a SSS charged black hole~\cite{Garfinkle:1990qj} which results from the coupling between a dilaton field and the Einstein-Maxwell theory.

As a final case, we would like to consider SSS metrics where the area function is general. A good example of such a case is the charged spherically symmetric dilaton black hole~\cite{Garfinkle:1990qj}. Here, in order to calculate the QLE, it is necessary to follow the same procedure as in Ref.~\cite{Chan:1996sx}, which generalizes the prescription of Brown York~\cite{Brown:1992br} to include theories of dilaton gravity.

This paper is organized as follows. In Section 2, we first review the Brown-York formalism. In particular, we apply it to spherically symmetric static black hole solutions, and also present the identity that relate the field energy with the gravitational charge at the horizon. In Section 3 we use the Brown-York formalism to find the QLE for regular black holes, and we derive an expression that relate the field energy with the gravitational charge in this case. Similarly, in Section 4 we evaluate the QLE for cases where $g_{00} \neq -(g_{11})^{-1}$, $g_{22} =-r^2$ and $g_{33} =- r^2 \sin \theta$, and we also derive an expression that relate the field energy with the gravitational charge.
In section 5 we present the results of the QLE of dilaton black hole solutions. Finally, Section 6 summarize our results. In the Appendix, we give the proof of the relation obtained in Section 4.

\section{Brown-York Quasilocal Energy for Spherically Symmetric Static Metrics}
\label{sec:2}
We consider the definition of QLE based on the covariant Hamilton-Jacobi formulation of general relativity,
proposed by Brown and York~\cite{Brown:1992br}
\begin{equation} E(r)=
\frac{1}{8 \pi}\int_B d^2x \sqrt{\sigma}(k-k_0) \,\,\label{B-Y} \, ,
\end{equation}
where $B$ is the two dimension spherical surface, $k$ is the trace of the extrinsic curvature of $B$,
$\sigma_{ij}$ is the metric of $B$ and $k_0$ is a reference term (for an asymptotically flat spacetime one choose Minkowski spacetime as the reference spacetime).

Let $ds^2$ be the line element for the most general static and spherically symmetric solution
\begin{equation} ds^2= -g(r) dt^2 + f(r)^{-1}dr^2 + R^2(r) d\Omega^2
\,\,\label{element} \, ,  \end{equation}
where $d\Omega^2 \equiv d\theta^2 + \sin^2\theta \, d\varphi^2$ and $f(r)$, $g(r)$ and $R(r)$ are arbitrary functions of the coordinate $r$.
If we choose the metric functions such that $g(r)=f(r)$ and the area function $R(r)=r$ with $f(r)=1-2m(r)/r$ and the boundary condition $m(r\rightarrow\infty)=M$ to satisfy asymptotic flatness, then the QLE inside a spherical surface of arbitrary radius $r\geq r_+$ associated to this line element is given by
\begin{equation}  E(r)= r-r\sqrt{1-\frac{2m(r)}{r}} \,\,\,
\,\,\label{energyBY} \, . \end{equation}
Note that if
$f(r_{\pm})=0$, then $E(r_{\pm})=r_{\pm}$, where $r=r_+$ and $r=r_-$ define the
surfaces called the outer and the inner horizons of
the black hole, respectively. The event horizon is at the radius $r_h=r_+$.

In addition, to obtain the QLE inside the horizon and particularly at the singularity, one could consider Ref.~\cite{Lundgren:2006fu} where a prescription for finding such values was presented. Thus the QLE inside the event horizon is
\begin{equation}  E(r)= r-r\sqrt{1-\frac{2m(r)}{r}} \,\,\, {\mbox{ if }}
\,\,\, r <r _{-}
\,\,\label{inside1} \,  \end{equation}
\begin{equation}  E(r)= r+r\sqrt{-1+\frac{2m(r)}{r}} \,\,\, {\mbox{ if }}
\,\,\, r_{-} < r <r_{+}
\,\,\label{inside2} \, . \end{equation}

As an application of the QLE, Bose and Dadhich in Ref.~\cite{Bose:1998uu} (see also Ref.~\cite{Dadhich:1997ze}) established an identity at the event horizon that relates the QLE to the Komar charge for spherically symmetric static black hole solutions by using the Gauss-Codazzi equations. These authors considered asymptotically flat solutions and other two particular cases: an asymptotically FRW black hole and a black hole with a global monopole charge. This identity is given by
\begin{equation}
E(r_h) - E(\infty) = M_H \,\,\label{B-D} \, ,
\end{equation}
where $M_H$ is the Komar charge evaluated at the horizon. Noting that the metric that we are considering has a single time-like Killing vector $\xi=\partial/\partial t$, the following relation holds
\begin{equation}
M_H=\kappa A/(4\pi) \,\,\label{Komar-H} \, ,
\end{equation}
where the surface gravity is $\kappa = f'(r_h)/2$ and the area of the sphere is $A=4\pi r_h^2$.

To illustrate the calculation of the QLE, we consider the known case of the Reissner-Nordstr\"{o}m metric mentioned in Ref.~\cite{Lundgren:2006fu}, where the line element is described by Eq.~(\ref{element}) with $R(r)=r$ and the metric functions are given by
\begin{equation}
f(r)=g(r)= 1 - \frac{2 M}{r} + \frac{Q^2}{r^2} \,\,\label{R-N} \, .
\end{equation}
In this case, if $M > |Q|$ is considered, the metric function $f(r)$ has two zeros, and therefore the black hole has two horizons which are located at $r_{\pm} = M \pm \sqrt{M^2 - Q^2}$, then we obtain $E(r_h)=r_h>E(\infty)=M$ and $dE(r)/dr<0$ from $r_+$ until $r\rightarrow\infty$. Similarly, as it was derived in Ref.~\cite{Lundgren:2006fu} applying Eq.~(\ref{inside1}), the evaluation of the QLE at the singularity is $E(0)=-|Q|$. This negative value was justified by the repulsive effect of the radial geodesics of massive neutral particles.

Moreover, we can establish whether the identity~(\ref{B-D}) is satisfied. Thus the gravitational field energy at the horizon is $E(r_h)-E(\infty)=\sqrt{M^2 - Q^2}$, and when we use Eq.~(\ref{Komar-H}), the Komar charge at the horizon becomes $M_H=\sqrt{M^2 - Q^2}$. Therefore, the relation~(\ref{B-D}) is clearly satisfied. If $M=|Q|$, we have an extremal black hole and $E(r_h)-E(\infty)=0$.

\section{Brown-York QLE for Regular Black Holes}
\label{sec:3}
We now compute the QLE and the Komar charge at the horizon of a nonlinear electrodynamics black hole and establish an identity of the type~(\ref{B-D}). To proceed, we write the identity~(\ref{B-D}) in terms of the mass function and its derivative.

We start by considering the following action of general relativity coupled to nonlinear electrodynamics
\begin{equation}
S = \frac{1}{16 \pi}\int d^4x \sqrt{-g} \left(R+\emph{L}(F)\right)
\,\,\label{action-NL} \, , \end{equation}
where $R$ is the scalar curvature, $g \equiv det|g_{\mu\nu}|$ and the Lagrangian $\emph{L}(F)$ is a nonlinear function of the Lorentz invariant $F=F^{\mu\nu}F_{\mu\nu}$. Likewise, the metric has the form of Eq.~(\ref{element}) with the mass function given by
\begin{equation}
m(r) = \frac{1}{4}\int L(F)r^2 dr + C
\,\,\label{m-function} \, , \end{equation}
where $C$ is an integration constant.

There are several regular black hole solutions that have been proposed in recent years. In order to study the relation given above, we consider the nonlinear electrodynamics coupled to gravity of Ref.~\cite{AyonBeato:1999rg}. Thus, according to this Ref. the mass function is defined as
\begin{equation}
m(r) = M - M\tanh\left(\frac{Q^2}{2 M r}\right)
\,\,\label{m-ABG} \, , \end{equation}
where $M$ and $Q$ are the mass and the charge of the black hole, respectively.
In this case, the condition  $|Q| < 1.0554 \,  M$ allows the outer and inner horizons. When $|Q| = 1.0554 \, M$ we have the extremal case. Note that if we evaluate the QLE when $|Q| \leq \, M$, its value approaches $M$ asymptotically from above of this value, that is, $E(r\rightarrow\infty)\rightarrow M^+$. On the other hand, if $|Q| > \, M$, its value approaches $M$ from below, that is,  $E(r\rightarrow\infty)\rightarrow M^-$. In this case, the type of behavior exhibited by the QLE can be understood by noting that in the large $r$ limit~\cite{Brown:1992br}
\begin{equation}  E(r)\approx M + \frac{1}{2 r}(M^2 - Q^2) \,\,\label{limit} \,  \end{equation}
because the metric function behaves asymptotically as $f(r)= 1-2M/r+Q^2/r^2+O(1/r^4)$.

If we return to Ref.~\cite{Lundgren:2006fu} and carry out a similar calculation using Eq.~(\ref{inside1}), we find that the QLE converges to $0$ as $r$ approaches $0$ for all $|Q| < 1.0554 \,  M$.

In this example of regular black hole, it is possible to show numerically that the inequality
\begin{equation}
E(r_h) - E(\infty) < M_H \,\,\label{ineq.B-D} \, ,
\end{equation}
is satisfied for any $|Q| < 1.0554 \,  M$.
As an illustration of this, we take the condition $|Q| \approx 1.048 \, M$, finding $E(r_h)=E(\infty)=M$, while $M_H\neq 0$.

The introduction of mass function helps us to clarify such an inequality. An equivalent identity to (\ref{B-D}) can be obtained by noting that $m(r_h)=r_h/2$, $m(\infty)=M$ and that $M_H=m(r_h)-r_h \, m'(r_h)$. Thus, we can rewrite Eq.~(\ref{B-D}) as
\begin{equation}
\Delta(r_h)\equiv\frac{d(r \, m(r))}{dr}|_{r=\infty} -\frac{d(r \, m(r))}{dr}|_{r=r_h}=0\,\,\label{B-D.other} \, .
\end{equation}
In light of the last relation, it is now easy to see that any metric obeys the identity~(\ref{B-D}) if the value of $(r m(r))'$ is a constant. In a similar way, one may see that if $(r m(r))'$ depends on the coordinate radius $r$, then such a metric clearly do not satisfy the identity~(\ref{B-D}). To illustrate this, let us consider again the example of Reissner-Nordstr\"{o}m. It is easily seen that the derivative of $m(r) \, r = r \, M - Q^2/2$ is in fact a constant. If we now consider the regular black hole where $m(r) \, r = r \, M - r \, M \, \tanh\left(\frac{Q^2}{2 M r}\right)$,
we find immediately that $\Delta(r_h)\neq 0$. Similar conclusion is obtained when we check the cases considered in Refs.~\cite{AyonBeato:1998ub,AyonBeato:1999ec,AyonBeato:2000zs,Dymnikova:2004zc}.

We therefore can write the following identity for regular black holes
\begin{equation}
E(r_h) - E(\infty) = M_H - \Delta(r_h) \,\,\label{B-D.reg} \, .
\end{equation}

The above procedure to obtain relation~(\ref{B-D.reg}) can also be used for an arbitrary spherical boundary of constant radial coordinate $R>r_h$.
Using the definition of surface pressure $\mathcal{P}$ given in Ref.~\cite{Brown:1992br} that is expressed in terms of the metric function, i.e.,
\begin{equation}
\mathcal{P}=\frac{1}{8 \pi}\left(\frac{d}{dr}(\sqrt{f(r)} \, )+\frac{\sqrt{f(r)}}{r}-\frac{1}{r}\right)
\,\,\label{relation} \, , \end{equation}
one can write the following identity for regular black holes
\begin{equation}
E(R) - E(\infty) = 2 \mathcal{P} A(R) \sqrt{f(R)} - \Delta(R) \,\,\label{B-D.Gral} \, ,
\end{equation}
where $A(R) = 4 \pi R^2$ is the surface area and, in this case,
\begin{equation}
\Delta(R)=\int^\infty_{R} d\left(\frac{d(r m(r))}{dr}\right)
\,\,\label{rel-gral} \, . \end{equation}
Note that if $(r m(r))'$ is constant such as in the case of a Reissner-Nordstr\"{o}m metric, then $\Delta(R)=0$ in the relation~(\ref{B-D.Gral}). On the other hand, if $(r m(r))'$ leads as result a function depending on the coordinate radius $r$, then $\Delta(R) \neq 0$.

It is clear, from Eqs.~(\ref{B-D.reg}) and (\ref{B-D.Gral}), that
\begin{equation}
\mathcal{P} A(R) \sqrt{f(R)}\rightarrow T S \,\,\, {\mbox{ when }}
\,\,\, R\rightarrow r_h
\,\,\label{limit} \, ,
\end{equation}
where $T$ is the temperature on $r_h$ and $S = \pi r_h^2$ is the entropy.

\section{Cases with g(r)$\neq$f(r) and R(r)=r}
\label{sec:4}

We can generalize the previous analysis to find a relation for black hole solutions by considering $f(r)\neq g(r)$ and $R(r)=r$ in the general form of the line element given by Eq.~(\ref{element}). For it, we proceed by writing $g(r)=e^{2\delta(r)}f(r)$ with $f(r)=1-2m(r)/r$ and the boundary conditions $\delta(r\rightarrow\infty)=0$ and $m(r\rightarrow\infty)=M$ to satisfy asymptotic flatness. Thus, in this type of metric, the following identity is satisfied (details can be found in the Appendix)
\begin{equation}
E(r_h) - E(\infty) = M_H e^{-\delta(r_h)} - \Delta(r_h) \,\,\label{B-D.non} \, ,
\end{equation}
where $\Delta(r_h)$ is defined as in Eq.~(\ref{B-D.other}) and $\Delta(r_h)=0$ if we are considering gravity coupled to lineal electrodynamics. The Komar charge at the horizon becomes (see e.g. Ref.~\cite{Heusler:1994wa})
\begin{equation}
M_H =\frac{r_h^2}{2} g'(r_h)\sqrt{\frac{f(r_h)}{g(r_h)}} = \frac{r_h^2}{2} e^{\delta(r_h)}f'(r_h)\,\,\label{komar.N-S} \, ,
\end{equation}
and as in the previous examples, we find that $M_H = 2 T S$, where now the surface gravity and the area of the sphere are $\kappa = e^{\delta(r_h)} f'(r_h)/2$ and $A=4\pi r_h^2$, respectively.

As an illustration of this relation, we consider the SSS black hole with Skyrme and winding number $B=1$~\cite{luckock1986}, where the action of the Einstein-Skyrme has a similar form to the action given in Eq.~(\ref{action-NL}). But, in this case, $\emph{L}(F)$ is the Lagrangian for the Skyrme model, which is a nonlinear function of the profile function $F=F(r)$. Expanding the mass function around $r_h$, one may find that~\cite{Shiiki:2005pb}
\begin{equation}
m'(r_h)=\frac{\pi F_\pi}{a} \sin^2(F_h) \left(1+\frac{2 \sin^2(F_h)}{(a F_\pi r_h)^2}\right)
\,\,\label{m1} \, ,
\end{equation}
where $F_\pi$ is the pion decay constant, $a$ is a dimensionless parameter fixed experimentally, $F_h \equiv F(r_h)$, and $m(0)=0$ and $F(0)=\pi$ are the regularity conditions. Thus, one can show numerically that $\Delta(r_h) \neq 0$, and that the Eq.~(\ref{B-D.non}) is fulfilled.

An example of a black hole solution with $\Delta(r_h)=0$ is the GHS dilaton black hole given by~\cite{Garfinkle:1990qj}
\begin{equation}  ds^2 = -\frac{\left(1-\frac{2 M e^{\phi_0}}{r}\right)}{\left(1-\frac{Q^2 e^{3\phi_0}}{M r}\right)} dt^2 + \frac{1}{\left(1-\frac{2 M e^{\phi_0}}{r}\right)\left(1-\frac{Q^2 e^{3\phi_0}}{M r}\right)}dr^2 + r^2 d\Omega^2
\,\,\label{GHS} \, ,
\end{equation}
where $\phi_0$ is the asymptotic constant value of the dilaton field and $r_h=2 M e^{\phi_0}$ for $Q^2 < 2 e^{-2 \phi_0} M^2$. Here $m(r) = M e^{\phi_0} + Q^2 e^{3\phi_0}/(2 M) - Q^2 e^{4\phi_0}/r$ and therefore
\begin{equation}
E(r_h) - E(\infty) = 2 M e^{\phi_0} - \left(M e^{\phi_0} + \frac{Q^2 e^{3\phi_0}}{2 M}\right)
\,\,\label{ex.ghs-1} \, ,
\end{equation}
which is the same as
\begin{eqnarray}
M_H  e^{-\delta(r_h)} &=& 2 \, T \, S \, e^{-\delta(r_h)}\nonumber \\ &=& 2 \, (\frac{1}{8 \pi M e^{\phi_0}}) \, (4 \pi M^2 e^{2\phi_0}) \left(1-\frac{Q^2 e^{3\phi_0}}{M r_h}\right)\nonumber \\
&=&M e^{\phi_0} \left(1-\frac{Q^2 e^{3\phi_0}}{M r_h}\right)
\,\,\label{ex.ghs-2} \, .
\end{eqnarray}
Thus, the GHS black hole obeys the identity~(\ref{B-D.non}) with $\Delta(r_h)=0$.

\section{Case with R(r)$\neq$r for dilaton gravity}
\label{sec:5}

In order to study a case of SSS metric for a more general radial function $R(r)$ it can be useful to consider a black hole solution to dilaton theory of gravity. In particular, we are interested in black hole solutions with $g(r)=f(r)$, $R(r)\neq r$, $f(r)=1-2m(r)/r$ in the line element given by Eq.~(\ref{element}) and the boundary condition at spatial infinity satisfies asymptotic flatness. Thus we can follow the analysis given in Ref.~\cite{Chan:1996sx} which studies the black holes in dilaton gravity and where the QLE is given by
\begin{equation} E(r)=-\frac{\sqrt{f(r)}}{2}\frac{d}{dr}(r^2 P^2(r))-E_0
\,\,\label{E-R} \, , \end{equation}
where $E_0$ is a reference energy.
Then, one may verify that the following identity is satisfied
\begin{equation}
E(r_h) - E(\infty) = M_H - \Delta(r_h)\,\,\label{B-D.R} \, ,
\end{equation}
where now the gravitational charge is given by
\begin{equation}
M_H=\frac{R^2(r_h)}{2}  f'(r_h)
\,\,\label{komar-R} \, ,
\end{equation}
and
\begin{equation}
\Delta(r_h)=\int^\infty_{r_h} d\left(\frac{d}{dr}(P^2(r) \, r \, m(r))\right)
\,\,\label{rel-R} \, ,
\end{equation}
here we have introduced $R(r)=r P(r)$. Note that $\Delta(r_h)=0$ if we consider lineal electrodynamics and $P(r)$ is constant. The Garfinkle-Horne dilaton black hole~\cite{Garfinkle:1990qj} is an example for which the relation~(\ref{B-D.R}) together with Eqs.~(\ref{komar-R}) and~(\ref{rel-R}) is applicable.

The formalism can also be used to calculate the QLE of the Schwarzschild black hole with a global monopole~\cite{Barriola:1989hx}, defined by $g(r)=f(r)=1-\eta^2-2m/r$ and $R(r)=r$, where $\eta$ is the charge of global monopole and $m$ is the mass parameter.
In order to apply these relations, it is necessary to introduce the following coordinate transformation
\begin{equation}
t \rightarrow \sqrt{1-\eta^2} \,\, t \,\,  ,  \,\,\,\, r \rightarrow \sqrt{1-\eta^2} \,\, r
\,\,\label{trans.coord} \, ,
\end{equation}
and defining a new mass parameter
\begin{equation}  \tilde{m}=\frac{m}{(1-\eta^2)^{3/2}}
\,\,\label{new-mass} \, ,
\end{equation}
we can rewrite the line element of the black hole with global monopole as
\begin{equation}  ds^2= -\left(1-\frac{2 \tilde{m}}{r}\right) dt^2 + \left(1-\frac{2 \tilde{m}}{r}\right)^{-1}dr^2 + (1-\eta^2) r^2 d\Omega^2
\,\,\label{element-monop} \, . \end{equation}
We now see that $P(r)$ is a constant and that $\Delta(r_h)=0$. Applying Eq.~(\ref{E-R}) to such a metric leads
\begin{eqnarray} E(r_h) - E( \infty) &=& P^2(r_h) \, r_h - P^2(r \rightarrow \infty) \, \tilde{m}(r \rightarrow \infty) \nonumber  \\
&=& (1-\eta^2) (2 \tilde{m} - \tilde{m}) = \frac{m}{\sqrt{1-\eta^2}}
\,\,\label{ident-monop} \,  .\end{eqnarray}
On the other hand, using Eq.~(\ref{komar-R}), we obtain
\begin{equation} M_h = \frac{1}{2} R^2(r_h) \,\, \frac{d}{dr}\left(1-\frac{2 \tilde{m}}{r}\right)_{r=r_h} = (1-\eta^2) \,  \tilde{m}
\,\,\label{ident-monop2} \, . \end{equation}
The same result was obtained in Ref.~\cite{Bose:1998uu} based in the Hawking-Horowitz prescription~\cite{Hawking:1995fd}.

\section{Conclusion}
\label{sec:6}

Writing each quantity in terms of the mass function, we have proved that the identity given in Ref.~\cite{Bose:1998uu} , i.e.,
\begin{equation}
E(r_h) - E(\infty) = M_H \,\,\label{B-D-dis} \, ,
\end{equation}
must not only require that the black hole be SSS, but also it be coupled to usual Maxwell theory when we are considering a charged black hole. The identity can also be generalized to cases with $g_{00}\neq-(g_{11})^{-1}$ where the factor $e^{-\delta(r_h)}$ appears with the Komar mass ($\delta(r_h)$ is nonzero in general). The term $\Delta(r_h)$ appears in cases with $g_{00}\neq-(g_{11})^{-1}$ in the same way as in cases with $g_{00}=-(g_{11})^{-1}$. That is, the term $\Delta(r_h)$ is determined by terms of the purely nonlineal part of the respective nonlineal model coupled to the black hole and the difference of both $M_H$ and $E(r_h)$ may be offset by it. In both cases was considered $R(r)=r$ or $P(r)$ constant such as in the black hole with global monopole, after the coordinates transformation given by~(\ref{trans.coord}).

For the regular black hole~\cite{AyonBeato:1999rg} studied in Sec.~\ref{sec:3} and for the case given in Ref.~\cite{Dymnikova:2004zc} the energy inside of a sphere of radius $r$ can be less that the total energy $E(\infty)$. If we recall the interpretation of the QLE in the Newtonian approximation given in Ref.~\cite{Brown:1992br}, then the work necessary to assemble a spherical shell of such a radius $r$, mass $M$ and charge $Q$ of particles brought from infinity would be positive, which is different of the Reissner-Nordstr\"{o}m black hole where the required work is always negative (or zero in the extremal case). In turn, note that the field energy could be zero or negative, however the gravitational charge, which is related to the entropy and the temperature of the black hole, cannot be zero (unless the black hole is extremal) or negative. Nevertheless, there are other black holes coupled to nonlinear electrodynamics (for example the cases considered in Refs.~\cite{AyonBeato:1998ub,AyonBeato:1999ec,AyonBeato:2000zs}) where the field energy never become zero or negative, but the relation~(\ref{B-D.reg}) is still valid. Note also that the solutions of Refs.~\cite{AyonBeato:1999rg,Dymnikova:2004zc} permit $|Q| > M$ (where $E(r\rightarrow\infty)\rightarrow M^-$) and that the cases in Refs.~\cite{AyonBeato:1998ub,AyonBeato:1999ec,AyonBeato:2000zs} do not present black holes solutions when $|Q| > M$.

For SSS black hole with $g_{00}\neq-(g_{11})^{-1}$ we obtained the relation~(\ref{B-D.non}) where the Komar integral depends explicitly not only upon $g_{00}$ but also upon $g_{11}$, precisely because they are distinct.

Finally, making again good use of the proof of the identity where each quantity is written in terms of the mass function, we have extended the procedure to SSS black holes with a general radial function $R(r)$. In particular, we have considered SSS dilaton black holes, and therefore our derivation of the relation~(\ref{B-D.R}) with (\ref{komar-R}) and (\ref{rel-R}) is based on the quasilocal formalism developed in Ref.~\cite{Chan:1996sx}, where the thermodynamics of such black hole solutions is analyzed. Note that the expression for $\Delta(r_h)$ is nonzero still requiring the usual Maxwell theory because of the nonconstant term $P(r)$ present in $R(r)$.

As in Ref.~\cite{Bose:1998uu} we have a nonvariational identity relating quantities at the horizon and at infinity, that is, in a different way to the conventional formulation of the laws of black hole mechanics.

\section*{Acknowledgments}

The author wish to thank Ver\'onica Raspa for her useful suggestions on the manuscript. This research was supported by a postdoctoral fellowship of
$\mbox{CONICET}$.

\appendix
\section*{Appendix: Proof of the Relation~(\ref{B-D.non})}\label{proof}

We show that
\begin{equation}
E(r_h) - E(\infty) = M_H e^{-\delta(r_h)} - \Delta(r_h) \,\,\label{B-D.non.dem} \, .
\end{equation}
We proceed by supposing that the metric satisfies
\begin{equation}
E(r_h) - E(\infty) = M_H + \Phi(r_h)\,\,\label{a-dem} \, .
\end{equation}
This relation can be expressed in terms of $m(r)$
\begin{eqnarray}
r_h - M &=& \frac{r_h^2}{2} f'(r_h)e^{\delta(r_h)}+ \Phi(r_h)  \nonumber         \\
&=& \frac{r_h^2}{2} \frac{2[m(r_h) - r_h m'(r_h)]}{r_h^2}e^{\delta(r_h)} + \Phi(r_h)
\,\,\label{dem} \, ,
\end{eqnarray}
or equivalently,
\begin{equation}
2 m(r_h) - M = [m(r_h) - r_h m'(r_h)]e^{\delta(r_h)} + \Phi(r_h)
\,\,\label{dem.2} \, .
\end{equation}
The term $\Phi(r_h)$ is therefore given by the equation
\begin{eqnarray}
\Phi(r_h) &=& 2 m(r_h) - M  - [m(r_h) - r_h m'(r_h)]e^{\delta(r_h)} \nonumber         \\
&=& m(r_h) - m(\infty)  + r_h m'(r_h) \nonumber         \\ && + [-m(r_h) + r_h m'(r_h)](e^{\delta(r_h)}-1) \nonumber         \\
&=&  [-m(r_h) + r_h m'(r_h)](e^{\delta(r_h)}-1)-\Delta(r_h)
\,\,\label{dem.3} \,
\end{eqnarray}
which leads to
\begin{eqnarray}
&& E(r_h) - E(\infty)\nonumber         \\  &=& [m(r_h) - r_h m'(r_h)] e^{\delta(r_h)} \nonumber         \\ && + [-m(r_h) + r_h m'(r_h)](e^{\delta(r_h)}-1) - \Delta(r_h)  \nonumber         \\
&=& \{[m(r_h) - r_h m'(r_h)]e^{\delta(r_h)}\}e^{-\delta(r_h)} - \Delta(r_h)
\,\,\label{b-dem} \, .
\end{eqnarray}

%
%

\end{document}